\documentclass[a4paper]{PoS}
\usepackage{subfigure}

\title{Jet Substructure Variables with the SiFCC Detector at 100 TeV}

\ShortTitle{Jet Substructure Variables with the SiFCC Detector at 100 TeV}

\author{\speaker{C.-H Yeh}$^a$,S.V. Chekanov$^b$ ,A.V. Kotwal$^{c}$,J. Proudfoot$^{b}$,S. Sen$^{c}$,N.V. Tran$^{d}$,S.-S Yu$^{a}$\\     
     \llap{$^a$}Department of Physics and Center for High Energy and High Field Physics, National~Central~University\\
     Chung-Li, Taoyuan City 32001, Taiwan\\
     \llap{$^b$}HEP Division, Argonne National Laboratory\\
     9700 S. Cass Avenue, Argonne, IL 60439, USA\\
     \llap{$^c$}Department of Physics, Duke University\\
     Durham, NC 27708, USA\\
     \llap{$^d$}Fermi National Accelerator Laboratory\\
     Batavia, IL 6051, USA\\
     E-mail:  \email{a9510130375@gmail.com},
     \email{chekanov@anl.gov},
     \email{kotwal@phy.duke.edu},
     \email{proudfoot@anl.gov},
     \email{sourav.sen@duke.edu},
     \email{ntran@fnal.gov},
     \email{syu@phy.ncu.edu.tw}}

\abstract{
Future experiments beyond the LHC era
will measure high-momentum bosons ($W$, $Z$, $H$) and top quarks with strongly
collimated decay products that form hadronic jets.
This paper describes the studies  
of the performance of jet substructure variables using 
the Geant4 simulation of a detector designed for  high energy $pp$ collisions at a 100 TeV collider.  
The two-prong jets from $Z' \rightarrow WW$ and three-prong jets from $Z' \rightarrow t\bar{t}$ are compared with 
the background from light quark jets, assuming  $Z'$ masses in the range 5 -- 40 TeV.   
Our results indicate that the performance of jet-substructure 
reconstruction improves with reducing transverse cell sizes of a hadronic calorimeter 
from $\Delta \eta \times \Delta \phi = 0.087\times0.087$
to $0.022\times0.022$ in most cases.} 

\FullConference{The 39th International Conference on High Energy Physics (ICHEP2018)\\
		4-11 July, 2018\\
		Seoul, Korea}

\begin{document}

Future high-energy experiments, such as FCC-hh and SppC, 
will measure high-momentum bosons ($W$, $Z$, $H$) and top quarks 
with strongly collimated decay products that form hadronic jets. This leads to many
challenges for detector technologies.  In particular, reconstruction of jet substructure  variables for boosted jets with transverse
momentum above 10~TeV
requires appropriate  cell sizes of hadronic
calorimeters  (HCALs). In order to estimate 
transverse segmentation  
of HCALs for very boosted objects at future experiments, 
we used a FCC-like detector geometry,
a software based on Geant4 simulation and Monte Carlo event samples as described in~\cite{Chekanov:2016ppq}.
 
In this study we simulated the $Z'$ bosons with the mass of 5, 10, 20, and 40 TeV. These particles are forced to decay to two light-flavor jets ($q\bar{q}$) as background and $W W$ or $t\bar{t}$ as signal, where $W$($\rightarrow q\bar{q}$) and $t$($ \rightarrow  W^+\>b \rightarrow q\bar{q} b$) decay hadronically. Using different configurations of HCAL geometry, we draw the receiver operating characteristic (ROC) curves to quantify the detector performance and find out the cell size that can give the best separation power to distinguish signal from background for different jet substructures.

\begin{figure}
 \begin{center}
   \begin{minipage}[c]{0.33\textwidth}
     \includegraphics[width=\textwidth]{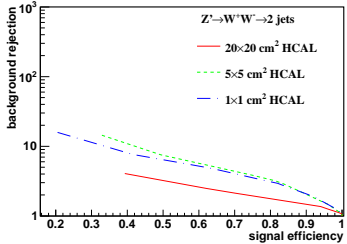}
   \end{minipage}\hfill
   \begin{minipage}[c]{0.6\textwidth}
     \caption{
        ROC curves for the $\tau_{21}$ substructure variable \cite{Thaler:2010tr} for the decay $Z'\rightarrow WW$ with a $Z'$ mass of 20 TeV. The calculations were performed for a HCAL with 
different transverse segmentations.
     } \label{1}
   \end{minipage}
    \end{center}
\end{figure}

We used several jet substructure variables, including jet soft-drop mass~\cite{Larkoski:2014wba}, $N$-subjettiness~\cite{Thaler:2010tr} and energy correlation function~\cite{Larkoski:2013eya} in this study. The variables considered are $\tau_{21}$ and $C_2^1$ for the $Z'\rightarrow WW$ signal and $\tau_{32}$ for the $Z' \rightarrow t\bar{t}$ signal. Figure~\ref{1} shows the ROC curves for $\tau_{21}$~\cite{Thaler:2010tr} using three HCAL cell sizes for jets with $P_{T}$$\sim$10 TeV. For the $\tau_{21}$ and $C_2^1$ variables, the $\Delta \eta \times \Delta \phi = 0.022\times0.022$ (or $5\times5$ cm$^2$) geometry gives optimal performance, while for the jet soft-drop mass variable,  the $\Delta \eta \times \Delta \phi = 0.0043\times0.0043$ (or $1\times1~\mathrm{cm}^2$) geometry gives optimal performance (followed by $5\times5$ cm$^2$). The results for the $\tau_{32}$ variable are inconclusive. 

In conclusion, the  performance 
of a  HCAL with 
$5\times5$ cm$^2$ cells is, in most cases,
better than for $\Delta \eta \times \Delta \phi$ = $0.087\times0.087$ (or $20\times20$ cm$^2$) cells.
Therefore, this study confirms the  baseline SiFCC detector geometry \cite{Chekanov:2016ppq}
that uses $5\times5$ cm$^2$ HCAL cells.

\end{document}